\documentstyle[12pt]{article}
\def\be{\begin{equation}}
\def\ee{\end{equation}}
\def\bea{\begin{eqnarray}}
\def\eea{\end{eqnarray}}

\begin{document}

\begin{center}
{\Large{\bf Rotating-Moving D-Branes with Background
Fields in the Superstring Theory}}

\vskip .5cm {\large Farzin Safarzadeh-Maleki and Davoud Kamani}
\vskip .1cm
{\it Department of Physics, Amirkabir University
of Technology \\
(Tehran Polytechnic)\\
P.O.Box: 15875-4413, Tehran, Iran}\\
{\sl e-mails: kamani@aut.ac.ir  ,  f.safarzadeh@aut.ac.ir}\\
\end{center}

\begin{abstract}

Using the boundary state formalism we study rotating and
moving D$p$-branes in the presence of the following background
fields: Kalb-Ramond, $U(1)$ gauge potential and the tachyon field.
The rotation and motion are in the brane volumes. The interaction
amplitude of two D$p$-branes will be studied, and specially
contribution of the superstring massless modes will be segregated. 
Because of the tachyon fields, rotations and velocities
of the branes, the behavior of the interaction amplitude reveals
obvious differences from what is conventional.

\end{abstract}



\vskip .5cm

\newpage

\section{Introduction}

D-branes as essential ingredients of the superstring theory 
\cite{1} have important applications in different aspects
of theoretical physics. These objects are classical solutions of 
the low-energy string effective action and hence can be 
described in terms of closed strings.
Besides, D-branes with nonzero background internal fields have 
shown several interesting properties \cite{2}-\cite{7}. For 
example, these fields affect the emitted closed strings of 
the branes and therefore modify the branes interactions.

On the other hand, we have the boundary state formalism 
for describing the D-branes \cite{8}-\cite{15} which is 
a useful tool in many complicated situations. This is
due to the fact that the boundary state encodes all relevant 
properties of the D-branes. Therefore, in the past years 
it has been widely used for studying properties of D-branes 
in the string theory. A boundary state can describe 
creation of closed string from vacuum, or equivalently it can 
be interpreted as a source for a closed string, emitted by a 
D-brane. Among achievements in this 
formalism it is its extension to the superstring theory and
considering the contribution of the conformal and super-conformal 
ghosts. The overlap of two boundary states corresponding to two D-branes, 
via the closed string propagator, gives the amplitude 
of interaction of the branes. So far this adequate method 
has been applied to the various configurations 
in the presence of different background fields. For instance,
some of these configurations are:
stationary branes, moving branes with constant 
velocities, angled branes \cite{16}-\cite{20}, various configurations
in the compact spacetime \cite{16}, in the presence 
of the tachyon field \cite{20}-\cite{21}, bound state of two D-branes 
\cite{14}, and so on. 

Previously we studied a general configuration of rotating and moving 
D$p$-branes of the bosonic string theory 
in the presence of the the following 
background fields: the Kalb-Ramond field, $U(1)$ gauge potentials  
which live in the D-branes worldvolumes and tachyon fields
\cite{21}. In this paper the same  
setup will be considered in the superstring theory. We shall 
see that the novelty of the results is considerable.
Our procedure is as follows. For this 
setup we obtain the boundary state,
associated with the brane, then we 
compute the interaction between 
two such D$p$-branes as a closed 
superstring tree-level diagram in the 
covariant formalism. The generality 
of the setup strongly recasts the 
feature of the boundary states and interaction of the branes. 
We shall observe that the interaction amplitude and its 
long-range part, which occurs between the distant branes, exhibit 
some appealing behaviors.  

Note that we shall consider rotation of each brane 
in its volume and its motion along the brane directions. Due to 
the various fields inside the brane there are preferred 
directions which indicate the breaking of the Lorentz 
symmetry and hence such rotation and motion are meaningful.

This paper is organized as follows. In Sec. 2, the boundary 
state of a closed superstring, corresponding to a rotating-moving 
D$p$-brane with various background fields will be constructed. 
In Sec. 3, interaction of 
two D$p$-branes in the NS-NS and R-R sectors of the superstring 
will be calculated. In Sec. 4, the long-range force of the 
interaction will be extracted. Section 5 is devoted to the 
conclusions.
\section{Boundary state associated with a rotating-moving 
D$p$-brane with background fields}

We use the following sigma-model action for closed string to 
describe a rotating and moving D$p$-brane, in the presence of the 
Kalb-Ramond, photonic and tachyonic fields 
\bea
S =&-&\frac{1}{4\pi\alpha'} {\int}_\Sigma
d^{2}\sigma(\sqrt{-g}g^{ab}G_{\mu\nu}\partial_a X^{\mu}\partial_b
X^{\nu}+\varepsilon^{ab} B_{\mu\nu}\partial_a X^{\mu}\partial_b
X^{\nu})
\nonumber\\
&+&\frac{1}{2\pi\alpha'} {\int}_{\partial\Sigma} d\sigma (
A_\alpha
\partial_{\sigma}X^{\alpha}+ \omega_{\alpha\beta}J^{\alpha\beta}_{\tau }
+T(X^\alpha)),
\eea
where $\Sigma$ is the worldsheet of the closed string, emitted (absorbed)
by the brane, and $\partial\Sigma$ shows the boundary of the worldsheet.
Besides, ``${\alpha }$'' and ``${\beta }$'' are indices along 
the brane worldvolume while ``${i}$'' will be used for the directions 
perpendicular to it. In addition, 
the background fields $G_{\mu \nu}$, $B_{\mu \nu}$, $A_{\alpha}$
and $T$, and also the antisymmetric variables 
$\omega_{\alpha \beta}$ and 
$J^{\alpha \beta}_\tau$ are the spacetime metric, 
Kalb-Ramond (an antisymmetric tensor), 
gauge field, tachyon field, angular 
velocity and angular momentum density 
of the brane, respectively.
Here we consider $G_{\mu \nu}$ as the flat spacetime metric with the
signature $\eta_{\mu\nu}= diag(-1, 1, \cdot\cdot\cdot, 1)$ and the
Kalb-Ramond field $B_{\mu \nu}$ to be a constant field.

In the presence of a D$p$-brane the 10-dimensional $U(1)$ 
gauge field $A_\mu$
is decomposed into a longitudinal $U(1)$ gauge field $A_\alpha$,
which lives in the worldvolume of the D$p$-brane, and a transverse 
part $A_i$ associated with the $9-p$ scalar fields, from the 
worldvolume point of view. 
These scalars represent coordinates of the brane. 
We shall keep them to be fixed,
that is, the branes do not have transverse motion. 
For the gauge field we choose the gauge
$A_{\alpha}=-\frac{1}{2}F_{\alpha \beta }X^{\beta}$ with the 
constant field strength. Now look at the tachyon. Usually in the
literature the tachyon field is nonzero just in one
dimension and its effects are studied on a space-filling
brane, while in the present article we consider a D$p$-brane 
with an arbitrary value for $p$. Besides, the square form of  
tachyon profile is used to produce a Gaussian integral, i.e.
$T(X)=\frac{1}{2}U_{\alpha\beta}X^{\alpha}X^{\beta}$ in which
the symmetric matrix $U_{\alpha \beta }$ is constant.
Thus, the tachyon field possesses components along all directions 
of the brane worldvolume. The gauge and tachyon fields are in 
the open string spectrum, which are attached to the D$p$-brane.
The brane's
rotation-motion term, that contains antisymmetric angular velocity
${\omega}_{\alpha \beta }$ and angular momentum density
$J^{\alpha \beta }_{\tau }$, is given by ${{\omega }_{\alpha \beta \
}J}^{\alpha \beta }_{\tau }=2{\omega }_{\alpha \beta }X^{\alpha
}{\partial }_{\tau }X^{\beta }$. In fact, the components
$\{\omega_{0 {\bar \alpha}}|{\bar \alpha} = 1, 2, \cdot\cdot\cdot,p\}$ 
denote the velocity of the brane, while the elements 
$\{\omega_{{\bar \alpha}{\bar \beta}}
|{\bar \alpha},{\bar \beta} = 1, 2, \cdot\cdot\cdot,p\}$ 
represent its rotation.
Note that in the presence of the antisymmetric field and the 
local gauge field there are preferred alignments in the brane,
and hence the rotation and motion of the brane in its
volume is sensible.
\subsection{Bosonic part of the boundary state}

In the closed string operator formalism the D-branes of the Type
IIA and Type IIB theories can be described by the boundary states. 
These are closed string states which insert a boundary on the closed 
string worldsheet and enforce on it appropriate boundary conditions.
Now we extract the corresponding boundary state for our setup. 
By vanishing of the variation of the action with respect to 
the closed string coordinates 
$X^{\mu }(\sigma ,\tau )$ the following boundary state equations 
are acquired
\bea
&~& [{(\eta }_{\alpha \beta }+4{\omega }_{\alpha \beta })
{\partial }_{\tau }X^{\beta }
+{{\mathcal F}}_{\alpha \beta}{\partial
}_{\sigma }X^{\beta } +U_{\alpha \beta }X^{\beta }]_{\tau =0}
|B_{\rm bos}\rangle=0 ,
\nonumber\\
&~& ({\delta X}^i)_{\tau =0}|B_{\rm bos}\rangle=0 ,
\eea
where ${\cal{F}}_{\alpha \beta}=\partial_\alpha A_\beta
-\partial_\beta A_\alpha - B_{\alpha \beta}$
is the total field strength. Note that we have assumed the 
following mixed elements vanish, i.e. 
$B_{\alpha i} =U_{\alpha i} =0 $.

It is worthwhile to show that along the worldvolume of 
the brane the Lorentz symmetry is broken. The Eqs. (2) leads to 
\bea 
J^{\alpha \beta}_{\rm bos}|B_{\rm bos}\rangle
= \int^\pi_0 d \sigma \bigg{[}
({\mathbf A}^{-1}{\cal{F}})^\alpha_{\;\;\;\gamma} X^\beta
\partial_\sigma X^\gamma
- ({\mathbf A}^{-1}{\cal{F}})^\beta_{\;\;\;\gamma} X^\alpha
\partial_\sigma X^\gamma
\nonumber\\
+ ({\mathbf A}^{-1}U)^\alpha_{\;\;\;\gamma} X^\beta X^\gamma
- ({\mathbf A}^{-1}U)^\beta_{\;\;\;\gamma} X^\alpha X^\gamma
\bigg{]}|B_{\rm bos}\rangle,
\eea
where ${\mathbf A}_{\alpha \beta}=\eta_{\alpha \beta}
+ 4\omega_{\alpha \beta}$. We observe that for restoring the Lorentz
invariance all elements of the tachyon matrix $U_{\alpha \beta}$
and the total field strength ${\cal{F}}_{\alpha \beta}$ must vanish. 
We demonstrated this for the bosonic part of the boundary state.
This procedure can also be applied for the total boundary state, 
which includes the bosonic and fermionic parts, 
to prove the breakdown of the Lorentz invariance along the 
worldvolume of the brane. 

Introducing the closed string mode expansion into Eq. (2) gives
\bea
&~& \left[\left({\eta }_{\alpha \beta }+4{\omega }_{\alpha \beta }
- {{\mathcal F}}_{{\mathbf \alpha }{\mathbf \beta }} +\frac{i}{2m}U_{\alpha
\beta }\right){\alpha }^{\beta }_m +{\left({\eta }_{\alpha \beta }+4{\omega
}_{\alpha \beta } +{{\mathcal F}}_{{\mathbf \alpha } {\mathbf
\beta }}\ -\frac{i}{2m}U_{\alpha \beta }\right)}{\widetilde{\alpha
}}^{\beta }_{-m}\right] {|B_{\rm bos}\rangle}^{\left({\rm osc}\right)}\ =0 ,
\nonumber\\
&~& \left [2 \alpha'{(\eta }_{\alpha \beta }+4{\omega
}_{\alpha \beta })p^{\beta } +U_{\alpha \beta }x^{\beta
}\right]{|B_{\rm bos}\rangle}^{\left(0\right)}\ =0,
\nonumber\\
&~& ({\alpha }^{i}_m-{\widetilde{\alpha }}^{i}_{-m})
{|B_{\rm bos}\rangle}^{\left({\rm osc}\right)}\ =0,
\nonumber\\
&~& (x^{i}-y^{i}){|B_{\rm bos}\rangle}^{\left(0\right)}\ =0,
\eea
where the set $\{y^i| i= p+1, \cdot \cdot \cdot, 9\}$ indicates 
the position of the brane. Besides, for the boundary state
$|B_{\rm bos} \rangle={|B_{\rm bos}\rangle}^{\left(0\right)}
\otimes {|B_{\rm bos}\rangle}^{\left({\rm osc}\right)}$
the components
${|B_{\rm bos}\rangle}^{\left(0\right)}$ and
${|B_{\rm bos}\rangle}^{\left({\rm osc}\right)}$ represent boundary
states for the zero modes and oscillating modes, respectively.

The solution of the oscillating part, which can be found by the coherent
state method, is given by
\bea
{|B_{\rm bos}\rangle}^{\left({\rm osc}\right)}\ 
=\prod^{\infty }_{n=1} {[\det
Q_{(n)}]^{-1}}\;{\exp \left[-\sum^{\infty }_{m=1}
{\frac{1}{m}{\alpha }^{\mu }_{-m}S_{(m)\mu \nu } {\widetilde{\alpha
}}^{\nu }_{-m}}\right]\ } {|0\rangle}_{\alpha}
\otimes {|0\rangle}_{\widetilde{\alpha }} \;,
\eea
where the matrices are defined as in the following
\bea
&~& Q_{(m){\alpha \beta }} = {\eta }_{\alpha \beta }+4{\omega
}_{\alpha \beta }-{{\mathcal F}}_{{\mathbf \alpha }{\mathbf \beta
}}+\frac{i}{2m}U_{\alpha \beta },
\nonumber\\
&~& S_{(m)\mu\nu}=\bigg{(}\frac{1}{2}\left[\Delta_{(m)}+
\left(\Delta^T_{(-m)}\right)^{-1}\right]_{\alpha \beta}
\; ,\; -{\delta}_{ij}\bigg{)},
\nonumber\\
&~& \Delta_{(m)\alpha \beta} = (Q_{(m)}^{-1}N_{(m)})_{\alpha
\beta},
\nonumber\\
&~& N_{(m){\alpha \beta }} = {\eta }_{\alpha \beta }+4{\omega
}_{\alpha \beta } +{{\mathcal F}}_{{\mathbf \alpha }{\mathbf
\beta }} -\frac{i}{2m}U_{\alpha \beta }.
\eea
Since the mode-dependent matrix $\Delta_{(m)}$ generally is not orthogonal
the matrix $\left(\Delta^T_{(-m)}\right)^{-1}$ also appears in the 
definition of $S_{(m)\mu\nu}$. In the Eq. (5) the normalization factor
$\prod^{\infty }_{n=1}{{[\det Q_{(n)}]}^{-1}}$
can be deduced from the disk partition function.

The boundary state for the zero modes finds the feature
\bea
{{\rm |}B_{\rm bos}\rangle}^{\left(0\right)}
&=& \int^{\infty }_{{\rm -}\infty }
\exp\left\{i{\alpha }^{{\rm '}}\left[\sum^{p}_{\alpha  =0}
{\left(U^{{\rm -}{\rm 1}}{\mathbf A}\right)}_{\alpha \alpha}
{\left(p^{\alpha}\right)}^{{\rm 2}}{\rm +}
\sum^{p}_{\alpha ,\beta {\rm =0},\alpha \ne \beta}{{\left(U^{{\rm -}{\rm
1}}{\mathbf A}+{\mathbf A}^T U^{-1}\right)}_{\alpha \beta }
p^{\alpha }p^{\beta}}\right]\right\}{\rm \ \ }
\nonumber\\
&\times& \left( \prod_{\alpha}{\rm |}p^{\alpha}\rangle dp^{\alpha}\right)
\otimes\prod_i{\delta {\rm (}x^i}{\rm -}y^i{\rm )}
{\rm |}p^i{\rm =0}\rangle .
\eea
The integration on the momenta indicates that the effects of
all values of the momentum components have been taken into account.
As we see, unlike the oscillating part, the total field 
strength did not entered in the Eq. (7). 

It should be noted that for calculating the interaction amplitude 
the contribution of the conformal ghosts $b$, $c$, ${\tilde b}$
and ${\tilde c}$ in the bosonic boundary state also will be taken 
into account.
\subsection{Fermionic part of the boundary state}

Since the supersymmetric version of the action (1) is invariant
under the global worldsheet supersymmetry, we can perform the 
supersymmetry transformations on the bosonic boundary
Eqs. (2) and transform them into their fermionic partners.
Therefore, one can use the following replacements
\bea
&~& {\partial }_+X^{\mu }\left(\sigma ,\tau
\right)\to -i\eta {\psi }^{\mu }_ +\left(\sigma ,\tau \right) ,
\nonumber\\
&~& {\partial }_-X^{\mu}\left(\sigma ,\tau \right)\to {\psi
}^{\mu }_-\left(\sigma ,\tau \right) ,
\eea
where $\eta = \pm 1$ has been introduced for the GSO projection
of the boundary state. As it was seen in the bosonic
boundary state equations, due to the presence of the tachyon field,
a replacement for
$X^{\mu }$ in terms of the fermionic components is also needed.
To obtain that, by using the replacements (8) and
${\partial}_{\pm}=\frac{1}{2}(\partial_\tau \pm \partial_\sigma )$
and integration, we receive
\bea
X^{\mu }\left(\sigma ,\tau \right)  \to \sum_{k}{\frac{1}{2k}
\bigg{(}}i{\psi }^{\mu }_{k\ }\ e^{-2ik(\tau
-\sigma )}+\eta {\widetilde{\psi }}^{\mu }_{k\ }\
e^{-2ik(\tau +\sigma )} \bigg{)}.
\eea

Now by introducing the replacements (8) and (9) into the Eqs. (2),
for the closed string boundary at $\tau= 0$, we obtain 
\bea
&~& \left[{\left({\eta }_{\alpha \beta }+4{\omega
}_{\alpha \beta } -{{\mathcal F}}_{{\mathbf \alpha } {\mathbf
\beta }}\ +\frac{i}{2k}U_{\alpha \beta }\right)}\psi^\beta_k
-i\eta \left({\eta }_{\alpha \beta }+4{\omega }_{\alpha \beta }+ {{\mathcal
F}}_{{\mathbf \alpha }{\mathbf \beta }} -\frac{i}{2k}U_{\alpha
\beta }\right){\widetilde \psi}^\beta_{-k}\right]
{|B^{{(\rm osc)}}_{\rm ferm} \;, \eta \rangle}\ =0 ,
\nonumber\\
&~& ({\psi}^{i}_{k}+i\eta {\widetilde{\psi }}^{i}_{-k})
{|B^{({\rm osc})}_{\rm ferm}\;,\eta\rangle} =0 ,
\eea
for the oscillating parts of the R-R and NS-NS sectors, and
\bea
&~& [({\eta }_{\alpha \beta }+4{\omega
}_{\alpha \beta } -{{\mathcal F}}_{{\mathbf \alpha } {\mathbf
\beta }})\psi^\beta_0
-i\eta ({\eta }_{\alpha \beta }+4{\omega }_{\alpha \beta }+ {{\mathcal
F}}_{{\mathbf \alpha }{\mathbf \beta }} ){\widetilde \psi}^\beta_0]
|B,\eta \rangle^{(0)}_{\rm R}=0 ,
\nonumber\\
&~& ({\psi}^i_0 +i\eta {\widetilde \psi}^i_0)
|B,\eta \rangle^{(0)}_{\rm R} =0 ,
\eea
for the zero-mode part of the R-R sector. As we see in this 
sector the tachyon has been omitted from the zero-mode boundary state.
The importance of this portion will be revealed in the R-R
sector of the boundary state.
The Eqs. (10) and (11) can be rewritten in the following features  
\bea
({\psi}^{\mu}_{k}-i\eta \ {S}^{\mu }_{(k)\;\;\nu}\ {\widetilde{\psi }}^{\nu
}_{-k})|B^{({\rm osc})}_{\rm ferm}\;,\eta \rangle=0,
\eea
for oscillating parts of both sectors, and
\bea
(d^{\mu}_0-i\eta \ {\bar S}^{\mu }_{\;\;\;\nu}\ {\widetilde d}^{\nu
}_0)|B,\eta \rangle^{(0)}_{\rm R}=0,
\eea
for the zero-mode part of the R-R sector. The matrix 
${\bar S}^{\mu }_{\;\;\;\nu}$ is defined by
\bea
&~& {\bar S}_{\mu\nu}=({\bar \Delta}_{\alpha \beta}\; ,\; -{\delta}_{ij}),
\nonumber\\
&~& {\bar \Delta}_{\alpha \beta} = ({\bar Q}^{-1}{\bar N})_{\alpha
\beta},
\nonumber\\
&~& {\bar Q}_{\alpha \beta } = {\eta }_{\alpha \beta }+4{\omega
}_{\alpha \beta }-{{\mathcal F}}_{{\mathbf \alpha }{\mathbf \beta
}},
\nonumber\\
&~& {\bar N}_{{\alpha \beta }} = {\eta }_{\alpha \beta }+4{\omega
}_{\alpha \beta } +{{\mathcal F}}_{{\mathbf \alpha }{\mathbf
\beta }}.
\eea

Note that in the fermionic parts we should also consider the boundary states 
associated with the super-conformal ghosts which will be needed for 
calculating the interaction amplitude.
\subsubsection{The Neveu-Schwarz sector}

Similar to the bosonic section, with the help of the coherent state
method, the oscillating part of the fermionic boundary state including 
both sectors can be calculated. Thus, the Eq. (12) implies that the 
NS-NS sector boundary state has the form
\bea
|B_{\rm ferm} , \eta \rangle_{\rm NS}=\prod^{\infty}_{r=1/2}[\det
Q_{(r)}]\exp \bigg{[}i\eta \sum^{\infty}_{r=1/2}(b^{\mu }_{-r}
S_{(r)\mu \nu}{\widetilde b}^{\nu}_{-r})\bigg{]}|0 \rangle_{\rm NS} .
\eea
When the path integral is computed the determinant is reversed
in comparing to the bosonic Eq. (5). This is due to the
Grassmannian property of the fermionic variables \cite{8}.
\subsubsection{The Ramond-Ramond sector}

Solving the Eqs. (12) and (13) in the R-R sector yields the following
boundary state
\bea
|B_{\rm ferm} \;, \eta \rangle_{\rm R}
=\prod^{\infty }_{n=1}[\det Q_{(n)}] {\exp  \left[i\eta
\sum^{\infty }_{m=1}{(d^{\mu }_{-m}S_{(m)\mu \nu }
{\widetilde{d}}^{\nu }_{-m})} \right]\ }
|B,\eta \rangle^{(0)}_{\rm R}.
\eea
The explicit form of the zero-mode state both in the Type IIA and Type IIB
theories is
\bea
|B,\eta \rangle^{(0)}_{\rm R}=
\bigg{[}C\Gamma^0 \Gamma^1 \cdot\cdot\cdot \Gamma^p
\bigg{(}\frac{1+i\eta\Gamma_{11}}
{1+i\eta}\bigg{)}\Omega\bigg{]}_{AB}|A\rangle \otimes
|\widetilde{B}\rangle,
\eea
where $A$ and $B$ denote the 32-dimensional indices 
for the spinors and $\Gamma$-matrices in the 10-dimensional spacetime, 
$|A\rangle \otimes|\widetilde{B}\rangle$ is the vacuum of the zero
modes $d^{\mu}_{0}$ and $\widetilde{d}^{\mu}_{0}\;$, $C$ is the charge 
conjugate matrix, and 
\bea
&~& \Omega= * \exp \left( \frac{1}{2}\Phi_{\alpha \beta}\Gamma^{\alpha}
\Gamma^{\beta}\right)*,
\nonumber\\
&~& \Phi_{\alpha \beta}=\bigg{(}({\bar \Delta}-1)
({\bar \Delta}+1)^{-1}\bigg{)}_{\alpha \beta}.
\eea
The notation *  * implies that one should expand the exponential 
and then antisymmetrize the indices of the $\Gamma$-matrices.
Therefore, since all terms in the expansion with repeated Lorentz 
indices are dropped, there are a finite number of terms for 
each value of $p$. As an example, for the D3-brane the matrix 
$\Omega$ takes the form 
\bea
\Omega= 1+\frac{1}{2}\sum^3_{\alpha , \beta =0}
\Phi_{\alpha \beta}\Gamma^{\alpha}\Gamma^{\beta}+
(\Phi_{01}\Phi_{23}-
\Phi_{02}\Phi_{13}+
\Phi_{03}\Phi_{12})
\Gamma^{0}\Gamma^{1}\Gamma^{2}\Gamma^{3}.
\nonumber
\eea

In fact, this convention implies that the matrix ${\bar \Delta}$ 
should be orthogonal which gives a restriction that the 
matrices $\omega$ and ${\cal{F}}$ should anticommute with each
other. For the D1-brane there is an electric field along the brane. 
Thus, according to this restriction, the
only element of the matrix $\omega$, i.e. the speed of the 
brane along itself, vanishes. This is an expected result, 
because of the direction of the electric field, motion
of the D-string along itself is not sensible. The other branes
can have both rotation and motion.
\section{Interaction of the branes}

Unbroken supersymmetry ensures that the Casimir energy of open 
superstrings is zero. Therefore, D-branes in supersymmetric 
configurations exert no net force on each other. 
A rotating/moving brane can break generically all the supersymmetries, 
and leads to orientation/velocity-dependent forces.

In this section we calculate the interaction of two
rotating and moving parallel D$p$-branes, equipped by background 
fields, via the closed string exchange. 
For both the NS-NS and R-R sectors the complete boundary state
can be written as the following product
\[|B,\eta \rangle_{\rm NS,R}=\frac{T_p}{2}|B_{\rm bos}\rangle
\otimes|B_{\rm gh}\rangle \otimes|B_{\rm ferm},\eta\rangle_{\rm NS,R}
\otimes|B_{\rm sgh},\eta\rangle_{\rm NS,R},\]
where the overall normalization factor $T_p$ is the D$p$-brane tension.
Note that the ghost and superghost boundary states are not affected by the 
rotation, motion and the background fields.
The explicit expressions of $|B_{\rm gh}\rangle$ and 
$|B_{\rm sgh}\rangle_{\rm NS,R}$ can be found in the 
literature, and hence we do not write them here.

For eliminating unwanted states, e.g. the closed string tachyon,
and in the same time making the number of
spacetime bosonic and fermionic physical excitations equal at
each mass level, as it is needed for supersymmetry, one should use the 
GSO projection. Therefore, the total boundary states which will be used
for calculation of the interaction find the forms
\bea
&~& |B \rangle_{\rm NS}=
\frac{1}{2}\left(|B,+\rangle_{\rm NS}
-|B,-\rangle_{\rm NS}\right),
\nonumber\\
&~& |B \rangle_{\rm R}=
\frac{1}{2}\left(|B,+\rangle_{\rm R}
+|B,-\rangle_{\rm R}\right).
\eea

One can obtain the interaction amplitude of two D-branes either 
by the open string one-loop or the closed string tree-level diagram.
Thus, the former is a quantum process while the latter is a 
classical process.
In the closed string picture the interaction between two D-branes 
is viewed as the exchange of a closed string between two
boundary states, geometrically describing a cylinder.
From this standpoint, the interaction is computed with a 
tree-level diagram. In this process a closed string is created
by one D-brane, it propagates in the transverse space between 
the two D-branes, and then the other D-brane absorbs it.
Therefore, the interaction amplitude 
between two D-branes in each sector is given by
the following overlap of the boundary states 
${\cal{A}}_{\rm NS-NS,R-R}\;=\;_{\rm NS,R}\langle
B_1|D |B_2\rangle_{\rm NS,R}$, where $D$ is
the closed string propagator. In other words, we have
\bea
{\cal{A}}_{\rm NS-NS,R-R}=2\alpha'\int^{\infty}_{0}dt\;
_{\rm NS,R}\langle B_1|e^{-tH_{\rm NS,R}}
|B_2\rangle_{\rm NS,R} .
\nonumber
\eea
The total closed superstring Hamiltonian $H_{\rm NS,R}$ is sum of 
the Hamiltonians of
the worldsheet bosons, fermions, conformal ghosts and super-conformal
ghosts in each sector. The complete interaction amplitude is given
by the following combination
\bea
{\cal{A}}_{\rm total}={\cal{A}}_{\rm NS-NS}+{\cal{A}}_{\rm R-R} .
\nonumber
\eea
According to this formula the boundary states are convenient 
tools for summing over all forces between two D-branes, which are 
mediated by the NS-NS and R-R states of closed superstring.
\subsection{The NS-NS sector interaction}

For maintaining the generality let's consider the $d$-dimensional 
spacetime instead of $d=10$.
Using the GSO projected boundary states (19) we obtain the
interaction amplitude, between two parallel D$p$-branes in the 
NS-NS sector, as follows
\bea
{{\mathcal A}}_{\rm NS-NS}&=&\frac{T_{p}^{2}V_{p+1}{\alpha }^{{\rm '\
}}}{8(2\pi)^{d-p-1}}
\prod^{\infty}_{m=1}\frac{\det[Q^\dagger_{(m-1/2)1}Q_{(m-1/2)2}]}
{\det \left[Q^\dagger_{\left(m\right)1}
Q_{\left(m\right)2}\right]}\
\nonumber\\
& \times &\int^{\infty }_0\ dt \bigg{\{}\frac{1}{\sqrt{\det
(R^\dagger_1 R_2)}} {\left(\sqrt{\frac{\pi } {{\alpha }^{{\rm '\
}}t}}\right)}^{d-p-1}{\exp\left(-\frac{1}{{4\alpha }^{{\rm '\
}}t}\sum_i{{\left(y^i_2-y^i_1\right)}^2}\right)}
\nonumber\\
&\times&\frac{1}{q}\bigg{(}\prod^{\infty
}_{n=1}{\left[{\left(\frac{1-q^{2n}}{1+q^{2n-1}}\right)}^{3+p-d}\;
\frac{\det ({\mathbf 1}+{H}^\dagger_{(n)1}H_{(n)2}\;q^{2n-1})\ }{\det
({\mathbf 1}-{H}^\dagger_{(n)1}H_{(n)2}\;q^{2n})}\right]}\
\nonumber\\
&-&\prod^{\infty}_{n=1}{\left[{\left(\frac{1-q^{2n}}{1-q^{2n-1}}
\right)}^{3+p-d}\;
\frac{\det ({\mathbf 1}-{H}^\dagger_{(n)1}H_{(n)2}\;q^{2n-1})\ }{\det
({\mathbf 1}-{H}^\dagger_{(n)1}H_{(n)2}\;q^{2n})}\right]}\bigg{)}\bigg{\}} ,
\eea
where the indices ``1'' and ``2'' refer to the first brane or 
$|B_1 \rangle$ and the second brane or $|B_2 \rangle$,
$V_{p+1}$ is the common worldvolume of the two D$p$-branes, 
$q=e^{-2t}$, $H_{(n)a}=(\Delta_{(n)a}+[\Delta^{-1}_{(-n)a}]^T)/2$ with  
$a=1,2 $, and the symmetric matrices $R_1$ and $R_2$ contain nonzero 
elements only along the branes worldvolumes
\bea
&~& (R_a)_{\alpha \beta }=2\alpha'(-i{\mathcal M}_a
-iU^{-1}_a{\mathbf A}_a -i{\mathbf A}^T_a U^{-1}_a
+t{\mathbf{1}})_{\alpha \beta },\;\;a=1,2,
\nonumber\\
&~& {\mathcal M}_a{\mathbf =}\left( \begin{array}{ccc}
{\left(U^{-1}_a{\mathbf A}_a\right)}_{00} & \cdots  & {\mathbf 0}\\
\vdots  & \ddots  & \vdots  \\
{\mathbf 0} & \cdots  & {\left(U^{-1}_a{\mathbf A}_a\right)}_{pp}
\end{array} \right),
\nonumber\\
&~& ({\mathbf A}_a)_{\alpha \beta }{\mathbf =}\eta_{\alpha \beta }
+4({\mathbf \omega }_a)_{\alpha \beta }.
\eea
In addition, we applied the relations $\langle p^\alpha | p^\beta \rangle
=2\pi \delta (p^\alpha - p^\beta)$ and $(2\pi)^{p+1} \delta^{(p+1)}(0)
=V_{p+1}$.

In this amplitude the exponential is a damping factor with
respect to the distance of the branes. In the last two products:
the determinant in the denominators reflects the portion of the bosons
oscillators along the branes worldvolumes, the determinants in the 
numerators are due to the fermions
oscillators again along the branes worldvolumes. The other 
factors in the products are contributions of the bosons and 
fermions oscillators, perpendicular to the brane worldvolume,
and also of the conformal ghosts and super-conformal ghosts. 
Explicitly, the power $3+p-d =2 - (d-p-1)$ is decomposed as: 
2 in the numerators for the ghosts, 2 in the denominators
for the superghosts,
$-(d-p-1)$ in the numerators for transverse oscillators
of the bosons and $-(d-p-1)$ in the denominators for transverse 
oscillators of the fermions. The remaining part of the integrand of 
the amplitude is overlap of the boundary states of the bosonic zero modes, 
i.e. the Eq. (7). This part completely is influenced by  
the internal tachyon fields, the motion and rotation of the branes.

Contributions of all closed superstring states in the NS-NS sector 
that the two branes can emit, are gathered in the amplitude (20). 
A part of the strength of the interaction is given by the constant 
overall factor of this amplitude, i.e. the first line of Eq. (20), 
which possesses contributions from the field parameters,
linear and angular velocities and the branes tensions.
\subsection{The R-R sector interaction}

Applying the total GSO projected boundary states (19) we acquire the
following interaction amplitude in the R-R sector
\bea
{{\mathcal A}}_{\rm R-R}&=&\frac{T_{p}^{2}V_{p+1}{\alpha
}^{{\rm '\ }}}{8(2\pi)^{d-p-1}}
\int^{\infty }_0{dt\ \bigg{\{} \left(\kappa \prod^{\infty
}_{n=1} {\left[{\left(\frac{1-q^{2n}}{1+q^{2n}}\right)}^{3+p-d}\;
\frac{\det ({\mathbf 1}+{H}^\dagger_{(n)1}H_{(n)2}q^{2n})\ }{\det
({\mathbf 1}-{H}^\dagger_{(n)1}H_{(n)2}q^{2n})\ }\right]} 
+{\kappa }^{{\rm '\
}}\right)}
\nonumber\\
&\times &\frac{1}{\sqrt{\det (R^\dagger_1 R_2)}}
{\left(\sqrt{\frac{\pi }{{\alpha }^{{\rm '\ }}t}}\right)}^{d-p-1}
{\exp\left(-\frac{1}{{4\alpha }^{{\rm '\
}}t}\sum_i{{\left(y^i_2-y^i_1\right)}^2}\right)}\bigg{\}},
\eea
where
\bea
&~& {\kappa }\equiv {
\frac{1}{2}(-1)^{p+1}\;{\rm Tr}[\Omega_{1}\;C^{-1}\;\Omega_{2}^T\;C]} ,
\nonumber\\
&~& {\kappa }^{{\rm '\ }}\equiv {i (-1)^p
\;{\rm Tr}[\Omega_{1}\;C^{-1}\;\Omega_{2}^T\;C\;\Gamma_{11}]} .
\eea
In above relations the matrices $\Omega_{1,2}$ have been defined 
by the Eq. (18) via the matrices
$\omega_{1,2}$ and ${\cal{F}}_{1,2}$ for the first and second branes.
As we can see in the R-R sector boundary state, and hence in the 
corresponding amplitude, the normalizing
determinant factors of the bosons and fermions cancel each other.

Now we are interested in the total amplitude, i.e. the combination
of the amplitudes in the NS-NS and R-R sectors.
In the total amplitude of the described system, the attraction
due to the exchange of the NS-NS states of closed string
is not compensated by the repulsion of the R-R states. Thus, we can
conclude that our setup does not satisfy the BPS no-force condition.
This is due to the fact that this configuration of the two 
D-branes does not preserve enough value of the spacetime 
supersymmetries of the Type IIA and Type IIB theories.
In fact, in the absence of the background fields, motions and 
rotations, the total amplitude vanishes, because this setup of 
the branes preserves half of the supersymmetry. 

A special feature of the non-BPS branes is presence of the 
tachyon field in their worldvolumes. In fact, it is not 
evident how the spacetime supersymmetry is realized with the 
tachyons, and existence of the broken supersymmetry in the 
presence of the tachyons has never been explicitly proven \cite{22}.
However, setting the branes in relative motion (or rotating them) 
breaks generically all the supersymmetries, and leads to 
velocity- or orientation-dependent forces \cite{23}.

We observe that in the amplitudes of both sectors, for  
a system of two D$(d-3)$-branes, the effect of the ghosts 
(superghosts) eliminates the contribution of the transverse 
oscillators of the bosons (fermions).
\subsection{An example}

To clarify our described system, let study a special case, i.e.
parallel $D2$-branes. Consider the $a$-th brane ($a=1,2$)
with the linear velocity $\{(v_{\bar \alpha})_a| {\bar \alpha}=1,2\}$,
the angular velocity $(\omega_{12})_a = \overline{\omega }_a$
and the fields
$(F_{0{\bar \alpha}})_a=(E_{\bar \alpha})_a$, $(F_{12})_a=B_{a}$
and $(U_{\alpha \beta})_a$. Therefore, the interaction
amplitude for the NS-NS sector is given by 
\bea
{{\mathcal A}}_{\rm NS-NS}&=&\frac{T_{2}^{2}V_3{\alpha }^{{\rm '\ }}}
{8(2\pi)^{d-3}}
\prod^{\infty}_{m=1}\frac{\det[Q^\dagger_{(m-1/2)1}Q_{(m-1/2)2}]}
{\det \left[Q^\dagger_{\left(m\right)1}
Q_{\left(m\right)2}\right]}\
\nonumber\\
&\times & \int^{\infty }_0{dt\ \bigg{\{} \  \frac{1}{\sqrt{\det
(R^\dagger_1 R_2)}} {\left(\sqrt{\frac{\pi } {{\alpha }^{{\rm '\
}}t}}\right)}^{d-3}{\exp\left(-\frac{1}{{4\alpha }^{{\rm '\
}}t}\sum_{i=3}^{d-1}{{\left(y^i_2-y^i_1\right)}^2}\right)}}
\nonumber\\
&\times & \frac{1}{q}\bigg{(}\prod^{\infty
}_{n=1}{\left[{\left(\frac{1-q^{2n}}{1+q^{2n+1}}\right)}^{5-d}\;
\frac{\det[({\mathbf 1}+H^\dagger_{(n)1}
H_{(n)2}q^{2n-1})]}{
\det[({\mathbf 1}-H^\dagger_{(n)1}
H_{(n)2}q^{2n})]}\right]}\
\nonumber\\
&-&\prod^{\infty}_{n=1}{\left[{\left(\frac{1-q^{2n}}
{1-q^{2n-1}}\right)}^{5-d}\; \frac{
\det[({\mathbf 1}-H^\dagger_{(n)1}
H_{(n)2}q^{2n-1})]}{
\det[({\mathbf 1}-H^\dagger_{(n)1}
H_{(n)2}q^{2n})]}\right]}\bigg{)}\bigg{\}},
\eea
where the matrix $H_{(n)a}$ is defined in terms of  
$Q_{(\pm n)a}$ and $N_{(\pm n)a}$, as before, in which
\bea
&~& Q_{(n)a}=\left( \begin{array}{ccc}
-1+\frac{iU_{00}}{2n} & 4v_1-E_1+\frac{iU_{01}}{2n}
& 4v_2-E_2+\frac{iU_{02}}{2n} \\
-4v_1+E_1+\frac{iU_{10}}{2n} & 1+\frac{iU_{11}}{2n}
& 4\overline{\omega }-B+\frac{iU_{12}}{2n} \\
-4v_2+E_2+\frac{iU_{20}}{2n} & -4\overline{\omega }+B+
\frac{iU_{21}}{2n} & 1+\frac{iU_{22}}{2n} \end{array}
\right)_a  \;,\;\; a=1\;,\;2,
\nonumber\\
&~& N_{(n)a}=\left( \begin{array}{ccc}
-1-\frac{iU_{00}}{2n} & 4v_1+E_1-\frac{iU_{01}}{2n}
& 4v_2+E_2-\frac{iU_{02}}{2n} \\
-4v_1-E_1-\frac{iU_{10}}{2n} & 1-\frac{iU_{11}}{2n}
& 4\overline{\omega }+B-\frac{iU_{12}}{2n} \\
-4v_2-E_2-\frac{iU_{20}}{2n} & -4\overline{\omega }
-B-\frac{iU_{21}}{2n} & 1-\frac{iU_{22}}{2n} \end{array}
\right)_a \; a=1\;,\;2.
\eea
The matrix elements of the symmetric matrix $R_a$ are as 
in the following 
\bea
&~& (R_a)_{00}=-2i{\alpha }^{{\rm '\ }}\
[\ (U^{-1})_{00}-4v_1(U^{-1})_{01}-4v_2(U^{-1})_{02}+it]_a\;,
\nonumber\\
&~& (R_a)_{01}=-2i{\alpha }^{{\rm '\ }}\
[\ 2(U^{-1})_{01}-4v_1\left((U^{-1})_{00}+(U^{-1})_{11}\right)
-4\overline{\omega }(U^{-1})_{02}-4v_2(U^{-1})_{21}+it\ ]_a\;,
\nonumber\\
&~& (R_a)_{02}=-2i{\alpha }^{{\rm '\ }}\
[\ 2(U^{-1})_{02}-4v_2\left((U^{-1})_{00}+(U^{-1})_{22}\right)
+4\overline{\omega }(U^{-1})_{01}-4v_1(U^{-1})_{12}+it\ ]_a\;,
\nonumber\\
&~& (R_a)_{11}=-2i{\alpha }^{{\rm '\ }}\
[\ (U^{-1})_{11}-4v_1(U^{-1})_{10}-4\overline{\omega }(U^{-1})_{12}+it]_a\;,
\nonumber\\
&~& (R_a)_{12}=-2i{\alpha }^{{\rm '\ }}\
[\ 2(U^{-1})_{12}+4\overline{\omega }\left((U^{-1})_{11}-(U^{-1})_{22}\right)
-4v_2(U^{-1})_{10}-4v_1(U^{-1})_{02}+it\ ]_a\;,
\nonumber\\
&~& (R_a)_{22}=-2i{\alpha }^{{\rm '\ }}\
[\ (U^{-1})_{22}-4v_2(U^{-1})_{20}+4\overline{\omega }(U^{-1})_{21}+it]_a ,
\eea
with $a=1\;,\;2$. Also, for the R-R sector the amplitude finds the feature
\bea
{{\mathcal A}}_{\rm R-R}&=&\frac{T_{2}^{2}V_3{\alpha }^{{\rm '\ }}}
{8(2\pi)^{d-3}}
\int^{\infty }_0dt\ \bigg{\{}
\frac{1}{\sqrt{\det (R^\dagger_1 R_2)}}
{\left(\sqrt{\frac{\pi }{{\alpha }^{{\rm '\ }}t}}\right)}^{d-3}
{\exp\left(-\frac{1}{{4\alpha }^{{\rm '\
}}t}\sum_{i=3}^{d-1}{{\left(y^i_2-y^i_1\right)}^2}\right)}
\nonumber\\
&\times & {\left(\kappa \prod^{\infty
}_{n=1} {\left[{\left(\frac{1-q^{2n}}{1+q^{2n}}\right)}^{5-d}\;
\frac{\det[({\mathbf 1}+H^\dagger_{(n)1}
H_{(n)2}q^{2n})]}{\det[({\mathbf 1}-H^\dagger_{(n)1}
H_{(n)2}q^{2n})]} \right]}+{\kappa }^{{\rm '\ }}\right)}\bigg{\}},
\eea
where
\bea
\kappa &=& 16 \left(-1+ \Phi_{(1)01}\Phi_{(2)01} +
\Phi_{(1)02}\Phi_{(2)02}-\Phi_{(1)12}\Phi_{(2)12}\right) ,
\nonumber\\
\kappa' &=& -\frac{1}{4} i\sum_{\alpha , \beta =0 }^2
\sum_{\alpha' , \beta' =0 }^2
\Phi_{(1)\alpha \beta}\Phi_{(2)\alpha' \beta'}{\rm Tr}
(\Gamma^\alpha \Gamma^\beta \Gamma^{\alpha'} \Gamma^{\beta'} 
\Gamma_{11}).
\eea
Note that we have used of $(\Gamma^\mu)^T=-C\Gamma^{\mu}C^{-1}$.
In fact, the D2-brane is the simplest brane which its rotation and
motion along its directions is sensible. We see that for this simple 
case the interaction amplitudes also are very complicated. 
\section{Interaction between distant D-branes}

For distant D-branes only the closed superstring massless states have a 
considerable contribution on the interaction. In other words, after 
long enough time, which is equivalent to the large distance of the 
branes, the massless states become dominant.
Technically, the contribution of these states on the interaction 
amplitude is obtained by taking the limit of the oscillators 
portions of Eqs. (20) and (22). 

Let $ P_{(n)} \in \{ -{\mathbf 1}\;,\;{\mathbf 1}\;,\;
-{H}^\dagger_{(n)1}H_{(n)2}\;,\;{H}^\dagger_{(n)1}H_{(n)2} \}$ 
and $q_n \in \{ q^{2n}\;,\;-q^{2n}\;,\;q^{2n-1} ,-\;q^{2n-1}\}$. 
By applying the following relation
\bea
\prod^{\infty}_{n=1}{\left({\det ({\mathbf 1}+q_n P_{(n)})\ }\right)} 
= {\exp  \
\left\{\sum^{\infty }_{k=0}{\left[\frac{{\left(-1\right)}^{k\
}}{k+1} \sum^{\infty
}_{n=1}{\rm Tr}(q_n P_{(n)})^{k+1}\right]}\right\}\ } ,
\eea
into the amplitudes (20) and (22) and sending $q$ to zero, the 
contribution of the massless states can be acquired. Therefore, 
{\it in the 10-dimensional spacetime}, we receive the following amplitudes
\bea
{\mathcal A}^{\rm (massless)}_{\rm NS-NS} &=&\frac{T_{p}^{2}
V_{p+1}{\alpha }^{{\rm '\
}}}{4(2\pi)^{9-p}}\prod^{\infty}_{m=1}\frac{\det[Q^\dagger_{(m-1/2)1}
Q_{(m-1/2)2}]}
{\det \left[Q^\dagger_{\left(m\right)1}
Q_{\left(m\right)2}\right]}
\int^{\infty } dt \bigg{\{} {\left( \sqrt{\frac{\pi }{\alpha' t}} 
\right)}^{9-p}
\nonumber\\
&\times &  \frac{[7-p +{\rm Tr}({H}^\dagger_{(1)1}H_{(1)2})]}
{\sqrt{\det (R^\dagger_1 R_2)}}\ \exp{\left(-\frac{1}{4\alpha't }
\sum_i {\left( y^2_i-y^1_i\right)}^2 \right)}\bigg{\}},
\eea
for the NS-NS sector, and
\bea
{\mathcal A}^{\rm (massless)}_{\rm R-R}
&=& \frac{T_{p}^{2}V_{p+1}{\alpha }^{{\rm '\
}}}{8(2\pi)^{9-p}}({\kappa }+{\kappa }^{{\rm '\ }})
\nonumber\\
& \times &\int^{\infty }{dt\ \bigg{\{} {\left(\sqrt{\frac{\pi }{{\alpha
}^{{\rm '\ }}t}}\right)}^{9-p}\frac{1}{\sqrt{\det (R^\dagger_1 R_2)}
}\ \exp\left(-\frac{1}{{4\alpha }^{{\rm '\
}}t}\sum_i{{\left(y^2_i-y^1_i\right)}^2}\right)}\bigg{\}},
\eea
for the R-R sector. We did not put the limit on the exponential
factors and the two other time dependent parts
$\left(\sqrt{\pi/(\alpha' t)}\right)^{9-p}$ and 
$1/\sqrt{\det(R^\dagger_1 R_2)}$ in the
Eqs. (30) and (31). The exponential
parts indicate the locations of the branes, while closed string
emission (absorption) does not depend on the positions of the branes.
The other two factors possess origin in the zero
modes, but not in the oscillators which define the closed
string states. The provenance of the factor 
$1/\sqrt{\det(R^\dagger_1 R_2)}$ is the tachyon fields which 
for large time weakens the interaction amplitudes. Precisely, 
since the presence of the open string tachyon makes the system unstable, 
after a long enough time the tachyon will roll down towards its 
minimum potential which causes a decreasing amplitude. In the 
absence of the tachyonic fields this slowing down factor disappears. 

We observe that 
for large distance branes the amplitude of the NS-NS sector 
depends on the total field strengths ${\cal{F}}_1$ and 
${\cal{F}}_2$ while these fields are absent in the R-R sector.
In other words, the internal electric and magnetic fields of
the branes impress the exchange of the graviton, dilaton and Kalb-Ramond 
states but do not modify the R-R repulsion force between the 
distant branes.

The total amplitude
\bea
{\mathcal A}^{\rm (massless)}
&=& {\mathcal A}^{\rm (massless)}_{\rm NS-NS}+
{\mathcal A}^{\rm (massless)}_{\rm R-R}
\nonumber\\
&=& \frac{1}{(\alpha') ^{3(p+1)/2}} \frac{{\bar T}_p^2 V_{p+1}}
{4 (2\pi)^{9-p}}\bigg{[} \frac{1}{2}(\kappa + \kappa')
+ \left( 7-p +{\rm Tr}({H}^\dagger_{(1)1}H_{(1)2})\right)
\nonumber\\
&\times& \prod^{\infty}_{m=1}\frac{\det \left( Q^\dagger_{(m-1/2)1}
Q_{(m-1/2)2}\right)} {\det \left( Q^\dagger_{\left(m\right)1}
Q_{\left(m\right)2}\right)} \bigg{]}
\nonumber\\
&\times& \int^{\infty} dt \bigg{\{} 
\left( \sqrt{\frac{\pi }{t}} \right)^{9-p}
\frac{1}{\sqrt{\det ({\bar R}^\dagger_1 {\bar R}_2)}}
\exp{\left( -\frac{L^2}{4 \alpha' t} \right)},
\eea
exhibits the long-range force between the D$p$-branes interaction,
where ${\bar T}_p = T_p|_{\alpha' =1}$,  
${\bar R}_{1,2} = R_{1,2}|_{\alpha' =1}$ and   
$ L^2 = \sum_i {\left( y^2_i-y^1_i\right)}^2 $
is the the square distance between the branes.
The NS-NS part indicates the exchange of the graviton, 
dilaton and Kalb-Ramond fields, in which the dilaton and the graviton 
give attraction force while the Kalb-Ramond gives repulsion one.
In the same way,
the R-R part indicates the repulsive contribution of the  
$(p+1)$-form potentials in the R-R sector. The net result force 
for the static branes with zero background fields vanishes, since 
the branes are BPS states. But when the branes possess velocity,
rotation and background fields the total force is nonzero, 
i.e. the various contributions are not balanced.
\section{Conclusions}

In this article we constructed a closed superstring boundary state
corresponding to a rotating and moving D$p$-brane which incorporates
configurations of electric, magnetic and tachyonic background fields.
The bosonic boundary state includes an exponential
factor which is absent in the conventional boundary states,
i.e. that one without tachyon.
This factor originates from the bosonic zero modes, rotation-motion
and tachyon terms in the boundary action. 

It should be mentioned that in
this article we considered the rotation axis perpendicular to
the branes. In addition, the branes move along their volumes.
According to the background fields we have preferred directions
in the branes which break the Lorentz invariance. Therefore,
these rotations and motions are meaningful.

According to the eigenvalues in the boundary state equations we deduce
the following constraint equation 
\bea
p^\alpha =-\frac{1}{2\alpha'}[ ( \eta +4\omega
)^{-1}U ]^\alpha_{\;\;\;\beta} x^\beta .
\nonumber
\eea
This implies that along the worldvolume of the brane, momentum of
an emitted (absorbed) closed string depends on its center of mass
position. Fountain of this relation completely is the tachyon field.
Thus, in the presence of the tachyon a closed string feels
an exotic potential which affects its evolution.

The boundary states enabled us to calculate the interaction
amplitude of two moving-rotating D$p$-branes with background fields.
This amplitude exponentially decreases with the square distance of
the branes, but it is a very complicated function of the setup
parameters. The variety of the adjustable parameters controls the
treatment of the interaction. For example, for two D$(d-3)$-branes,
which can have different background fields and different motions, 
the contribution of the (super-)ghosts removes the effects of all
transverse oscillators. It was shown that even for co-dimension
parallel branes with similar fields, the total amplitude
is nonzero. That is, our system does not satisfy the BPS no-force
condition. This is due to the presence of the rotations, velocities
and tachyonic and photonic fields on the branes. 

The long-range part of the interaction was extracted. In this
domain the instability of the branes, due to the background
tachyon fields, weakens the interaction. This decreasing behavior
can be understood by dissipation of the branes to the bulk modes
because of the rolling of the tachyon to its minimum potential
in long time regime. Finally, we observed that the internal electric 
and magnetic fields of the branes do not impress the R-R repulsion 
force between the large separated branes.

\end{document}